\documentclass[conference]{IEEEtran}
\usepackage{cite}
\usepackage[cmex10]{amsmath}
\usepackage{amssymb,amsfonts}
\usepackage{algorithmic}
\usepackage{algorithm}
\usepackage{graphicx}
\usepackage{textcomp}
\usepackage{xcolor}
\usepackage{array}
\usepackage{diagbox}
\usepackage{geometry}
\usepackage[caption=false,font=footnotesize]{subfig}

\newtheorem{remark}{Remark}

\def\BibTeX{{\rm B\kern-.05em{\sc i\kern-.025em b}\kern-.08em
    T\kern-.1667em\lower.7ex\hbox{E}\kern-.125emX}}
\begin{document}

\title{Low Complexity Iterative Detection for a Large-scale Distributed MIMO Prototyping System}

\author{{Yuan Feng$^1$, Menghan Wang$^1$, Dongming Wang$^1$, Xiaohu You$^1$}\\
$^1$National Mobile Comm. Research Lab., Southeast Univ., Nanjing, China; Email: wangdm@seu.edu.cn
}

\maketitle

\begin{abstract}
In this paper, we study the low-complexity iterative soft-input soft-output (SISO) detection algorithm in a large-scale distributed multiple-input multiple-output (MIMO) system. The uplink interference suppression matrix is designed to decompose the received multi-user signal into independent single-user receptions. An improved minimum-mean-square-error iterative soft decision interference cancellation (MMSE-ISDIC) based on eigenvalue decomposition (EVD-MMSE-ISDIC) is proposed to perform low-complexity detection of the decomposed signals. Furthermore, two iteration schemes are given to improve receiving performance, which are iterative detection and decoding (IDD) scheme and iterative detection (ID) scheme. While IDD utilizes the external information generated by the decoder for iterative detection, the output information of the detector is directly exploited with ID. In particular, a large-scale distributed MIMO prototyping system is introduced and a 32$ \times $32 (4 user equipments (UEs) and 4 remote antenna units (RAUs), each equipped with 8 antennas) experimental tests at 3.5 GHz was performed. The experimental results show that the proposed iterative receiver greatly outperforms the linear MMSE receiver, since it reduces the average number of error blocks of the system significantly. 
\end{abstract}

\IEEEpeerreviewmaketitle
\section{Introduction}
Large-scale distributed MIMO (D-MIMO) has attracted much attention due to its great potential to improve the spectral efficiency and the energy efficiency of wireless communication systems \cite{joung2014energy,you2010cooperative}. Compared with the conventional point-to-point MIMO system, large-scale D-MIMO system takes the advantages of both spatial multiplexing and macro-diversity \cite{almelah2017spectral,wang2013spectral}. Meanwhile, there are challenges in large-scale D-MIMO systems posed by several design aspects, such as channel estimation, hardware implementation, and detection complexity. In particular, a critical design challenge in such system is to design a reliable and computationally efficient detector even if the number of antennas grows very large and the complexity of joint processing is rather high \cite{elghariani2016low}. 

The performance and complexity of several detection algorithms in the reverse link for single-cell massive MIMO systems are compared in \cite{rusek2013scaling}, such as linear minimum-mean-square-error(LMMSE), tabu search (TS) and fixed complexity sphere decoder (FCSD). It was revealed that LMMSE receiver can operate very close to the optimal maximum-likelihood (ML) detector while saving an order of magnitude or more in the number of antennas. In order to further improve the performance of the receiver, MMSE-based iterative soft decision interference cancellation (MMSE-ISDIC) was proposed in \cite{rossler2003mmse,cao2007relation}, which is a iterative soft-input soft-output (SISO) detector and can be implemented in both uncoded and coded systems. However, it becomes noncompetitive when used to serve large-scale systems since it suffers from impractically high computational complexity.

We consider large-scale D-MIMO systems and our purpose is to develop efficient low-complexity iterative receivers for these systems. Firstly, the uplink interference suppression matrix is designed to decompose the received multi-user signal into independent single-user receptions. Then, a novel low-complexity solution named as MMSE-ISDIC based on eigenvalue decomposition (EVD-MMSE-ISDIC) is proposed to process the decomposed signals, which can reduce the complexity of the inversions of large-scale matrices via EVD. Two iteration schemes are given based on the low-complexity EVD-MMSE-ISDIC to improve the receiving performance of large-scale D-MIMO systems. While iterative detection and decoding (IDD) scheme utilizes the external information generated by the decoder as the \emph{a priori} information for iteration, the output information of the detector is directly exploited with iterative detection (ID) scheme to form iterative detector. In addition, the method of using the external information as the \emph{a priori} information to calculate the symbol mean and variance for iterations is discussed. Finally, a 32$ \times $32 large-scale D-MIMO prototyping system is introduced to validate the proposed iterative receiver.

The rest of this paper is organized as follows. Section \ref{System model} presents system configuration and channel model. In Section \ref{Uplink interference suppression}, we discuss the method of designing interference suppression matrix in large-scale D-MIMO system. In Section \ref{Iterative receiver based on evd-mmse-isdic}, the proposed EVD-MMSE-ISDIC algorithm is introduced. Section \ref{Large-scale D-MIMO prototyping system} describes the large-scale D-MIMO prototyping system and Section \ref{Experimental results} presents experimental results and related discussions. Finally, Section \ref{Conclusion} concludes the paper.

The following notation is used in the sequel. Vectors and matrices are denoted by lowercase and uppercase boldface letters, such as ${\bf{x}}$ and ${\bf{A}}$, respectively. $\left|  \cdot  \right|$ denotes the absolute value of a scalar. ${\left(  \cdot  \right)^{\rm{T}}}$ and ${\left(  \cdot  \right)^{\rm{H}}}$ represent transpose and Hermitian transpose, respectively. ${\text{diag}}\left( {\mathbf{x}} \right)$ represents a diagonal matrix with ${\mathbf{x}}$ along its diagonal. We use ${\cal CN}\left( {0,{\sigma ^2}} \right)$ to denote a circularly symmetric complex Gaussian variable with zero mean and covariance ${\sigma ^2}$.

\section{System model}\label{System model}
Consider a large-scale D-MIMO system with $M$ remote antenna units (RAUs) and $K$ users. ${N_{\rm{R}}}$ and ${N_{\rm{U}}}$ denote the number of antennas employed by RAU and user, respectively. All transmitted symbols are mapped from a QAM constellation $\Theta $. We assume that the number of data streams currently supported by the $k$th user is ${N_{{\rm{RI}},k}}$. For the $k$th user, the ${N_{{\rm{RI}},k}}$-length information sequence ${{\mathbf{s}}_k}{\text{ = }}{\left[ {{s_{k,1}},{s_{k,2}}, \ldots ,{s_{k,{N_{{\text{RI}},k}}}}} \right]^{\text{T}}}$  is fed to the precoder, where ${{\mathbf{s}}_k}$ is mapped to an ${N_{\rm{U}}}$-length sequence ${{\bf{s}}_k^{\prime}}$ by a precoding matrix ${{\mathbf{P}}_k} \in {\mathbb{C}^{{N_{\rm{U}}} \times {N_{{\rm{RI}},k}}}}$. The precoded sequence ${{\bf{s}}_k^{\prime}}$ is given as
\begin{equation}
{{\bf{s}}_k^{\prime}} = {{\bf{P}}_k}{{\bf{s}}_k}{\text{ = }}{\left[ {{{s}_{k,1}^{\prime}},{{s}_{k,2}^{\prime}}, \ldots ,{{s}_{k,{N_{\text{U}}}}^{\prime}}} \right]^{\text{T}}}.
\end{equation}
The received signal of RAUs is matched filtered, sampled and formed into 
${\mathbf{y}}{\text{ = }}{\left[ {{y_1},{y_2}, \ldots ,{y_{M{N_{\text{R}}}}}} \right]^{\text{T}}}$. 
Assuming adequate spatial separation and rich scattering between the RAU antenna elements, the received signal ${\bf{y}}$ 
can be expressed as 
\begin{equation}
{\mathbf{y}} = {\mathbf{Gs}^{\prime}}{\text{ + }}{\mathbf{n}},
\end{equation}
where ${\mathbf{s}^{\prime}} = {\left[ {{\mathbf{s}^{\prime}}_1^{\text{T}}, \ldots ,{\mathbf{s}^{\prime}}_k^{\text{T}}, \ldots ,{\mathbf{s}^{\prime}}_K^{\text{T}}} \right]^{\text{T}}}$ represents the signals sent by all users. 
We define
${\mathbf{G}} \triangleq \left[ {{{\mathbf{G}}_{\text{1}}}, \ldots ,{{\mathbf{G}}_k}, \ldots ,{{\mathbf{G}}_K}} \right]$ where 
${{\mathbf{G}}_k} \triangleq {\left[ {{{\mathbf{G}}_{1,k}}, \ldots ,{{\mathbf{G}}_{m,k}}, \ldots ,{{\mathbf{G}}_{M,k}}} \right]^{\text{T}}}$
with  
${{\mathbf{G}}_{m,k}} \in {\mathbb{C}^{{N_{\rm{R}}} \times {N_{\rm{U}}}}}$ 
represents the channel gain matrix from the $k$th user to the $m$th RAU. ${\bf{n}}$ is the complex additive white Gaussian noise (AWGN) with i.i.d. entries $\sim{\cal CN}\left( {0,{\sigma ^2}} \right)$.

\section{Uplink interference suppression}\label{Uplink interference suppression}
Since the RAUs receive signals from $K$ users simultaneously, it is necessary to perform interference suppression to eliminate inter-user interference. The interference suppression matrix ${{\mathbf{W}}_{{\text{IS}}}}$ should satisfy
\[{\bf{W}}_{{\rm{IS}}}{{\bf{G}}} = \left[ {\begin{array}{*{20}{c}}
{{\bf{R}}_1^{\rm{T}}}&{}&{}\\
{}& \ddots &{}\\
{}&{}&{{\bf{R}}_K^{\rm{T}}}
\end{array}} \right],\]
i.e., the product of the interference suppression matrix and the uplink channel is a block-diagonal matrix.
 Therefore, in order to obtain the desired ${{\mathbf{W}}_{{\text{IS}}}}$, singular value decomposition (SVD) on the channel matrix should be performed. For the $k$th user, we define 
${{\mathbf{G}}_{\left[ k \right]}} \triangleq \left[ {{{\mathbf{G}}_{\text{1}}}, \cdots ,{{\mathbf{G}}_{k{\text{ - }}1}},{{\mathbf{G}}_{k + 1}}, \cdots ,{{\mathbf{G}}_K}} \right]$, and it can be decomposed as 
\[{{\mathbf{G}}_{\left[ k \right]}} = {{\mathbf{U}}_k}{{\mathbf{\Lambda }}_k}{\mathbf{V}}_k^{\text{H}} = \left[ {{\mathbf{U}}_k^{(1)}{\mathbf{U}}_k^{(0)}} \right]\left[ {{\mathbf{\Lambda }}_k^{(1)}{\mathbf{0}}} \right]{\mathbf{V}}_k^{\text{H}},\]
where 
${{\mathbf{\Lambda }}_k}$
is diagonal matrix, in which the diagonal elements are singular values of 
${{\mathbf{G}}_{\left[ k \right]}}$ 
arranged in descending order. 
${\mathbf{\Lambda }}_k^{(1)}$ 
represents all non-zero singular values of 
${{\mathbf{G}}_{\left[ k \right]}}$. 
Assuming the rank of 
${{\mathbf{G}}_{\left[ k \right]}}$ 
is 
${r_k}$, 
${\mathbf{U}}_k^{(1)}$ 
contains the first 
${r_k}$
left-singular vectors and 
${\mathbf{U}}_k^{(0)}$ 
contains the remaining  
$\left( {{MN_{\text{R}}} - {r_k}} \right)$ 
left-singular vectors of ${{\mathbf{G}}_{\left[ k \right]}}$. Therefore,  
${\mathbf{U}}_k^{(0)}$ 
satisfies 
\[{\mathbf{G}}_k^{\text{H}}{\mathbf{U}}_k^{(0)}{\text{ = }}{{\mathbf{V}}_k}{\left( {{\mathbf{\Lambda }}_k^{(1)}} \right)^{\text{H}}}{\left( {{\mathbf{U}}_k^{(1)}} \right)^{\text{H}}}{\mathbf{U}}_k^{(0)}{\text{ = }}{\mathbf{0}}.\]
We choose ${N_{\text{R}}}$ rows of ${\left( {{\mathbf{U}}_k^{(0)}} \right)^{\text{H}}}$ randomly and denote it as ${{\mathbf{W}}_{{\text{IS}},k}}$. Thus, the uplink interference suppression matrix is derived as
${{\mathbf{W}}_{{\text{IS}}}}{\text{ = }}{\left[ {{{\mathbf{W}}_{{\text{IS}},1}^{\text{T}}}, \cdots ,{{\mathbf{W}}_{{\text{IS}},k}^{\text{T}}}, \cdots ,{{\mathbf{W}}_{{\text{IS}},K}^{\text{T}}}} \right]^{\text{T}}}$.

With interference suppression, the received multi-user signal is decomposed into $K$ independent single-user receptions, which can be expressed as
\begin{equation}
{{\bf{\tilde y}}_k} = {{\bf{\tilde H}}_k}{{\bf{s}}_k} + {{\bf{\tilde n}}_k},
\end{equation}
where
${{\mathbf{\tilde H}}_k} \triangleq {{\mathbf{W}}_{{\text{IS}},k}}{{\mathbf{H}}_k}$ 
and 
${{\mathbf{\tilde n}}_k} \triangleq {{\mathbf{W}}_{{\text{IS}},k}}\sum\limits_{l \ne k}^K {{{\mathbf{H}}_l}{{\mathbf{s}}_l}} {\text{ + }}{{\mathbf{W}}_{{\text{IS}},k}}{{\mathbf{n}}_k}$.  
${{\mathbf{H}}_k} \triangleq {\mathbf{GP}_k}$
is the equivalent channel matrix between RAUs and the $k$th user, which can be estimated by demodulation reference signal (DM-RS). ${\bf{n}_k}$ is the additive noise with i.i.d. entries $\sim{\cal CN}\left( {0,{\sigma ^2}} \right)$.

\section{Iterative receiver based on evd-mmse-isdic}\label{Iterative receiver based on evd-mmse-isdic}
In this section, we review the MMSE-ISDIC algorithm and propose the low complexity solution. In \ref{EVD-MMSE-ISDIC}, the improved algorithm named as EVD-MMSE-ISDIC is described. Then, we give two specific block-wise implementation, which are IDD scheme and ID scheme, based on EVD-MMSE-ISDIC to improve the receiving performance of large-scale D-MIMO systems. In \ref{Calculation of Symbol Mean and Variance}, the method of calculating the mean and variance from the \emph{a priori} information is discussed. Throughout the paper, the extrinsic messages and the \emph{a priori} information are in log-likelihood ratio (LLR) form.
\subsection{EVD-MMSE-ISDIC}\label{EVD-MMSE-ISDIC}
Consider the concatenation of detector and decoder. Utilizing the received signal ${{\mathbf{\tilde y}}_k}$, mean of received signal ${{\mathbf{\bar s}}_k}$ and variance of received signal ${{\mathbf{v}}_k}$, the traditional ISDIC output of the SISO detector is written as 
\begin{equation}
{{\bf{\hat s}}_k} = {{\bf{\bar s}}_k} + {{\bf{V}}_k}{\bf{\tilde H}}_k^{\rm{H}}{\left[ {{{{\bf{\tilde H}}}_k}{{\bf{V}}_k}{\bf{\tilde H}}_k^{\rm{H}} + {{\bf{\Sigma }}_k}} \right]^{ - 1}}\left[ {{{{\bf{\tilde y}}}_k} - {{{\bf{\tilde H}}}_k}{{{\bf{\bar s}}}_k}} \right],
\label{Bayesian}
\end{equation}
where ${{\mathbf{V}}_k} \triangleq {\text{diag}}\left( {{{\mathbf{v}}_k}} \right)$ and 
\begin{equation}
\begin{aligned} 
{{\mathbf{\Sigma }}_k} &\triangleq {\text{Cov}}\left( {{{{\mathbf{\tilde n}}}_k},{{{\mathbf{\tilde n}}}_k}} \right) \\ &= {{\mathbf{W}}_{{\text{IS}},k}}\left( {\sum\limits_{l \ne k}^K {{{\mathbf{H}}_l}{\mathbf{H}}_l^{\text{H}}} } \right){\mathbf{W}}_{{\text{IS}},k}^{\text{H}}{\text{ + }}{\sigma ^2}{{\mathbf{I}}_{{N_{\text{U}}}}}\label{formula1}
\end{aligned} 
\end{equation}
denotes the covariance matrix of ${{\bf{\tilde n}}_k}$.
Then, soft demodulation is performed on ${{\mathbf{\hat s}}_k}$ to generate extrinsic messages, which can be used as the \emph{a priori} information to calculate the mean and variance of ${{\mathbf{\hat s}}_k}$. It will be discussed in detail in section \ref{Calculation of Symbol Mean and Variance}. 

The updated ${{\mathbf{\bar s}}_k}$ and ${{\mathbf{v}}_k}$ are returned to the detector, which constructs the iterative receiver. It should be noted that the iteration is implemented in form of blocks, i.e., the vectors ${{\mathbf{\bar s}}_k}$, ${{\mathbf{v}}_k}$, ${{{\bf{\tilde y}}}_k}$ and ${{\mathbf{\hat s}}_k}$ are reconstructed into ${{\mathbf{\bar S}}_k^{\prime}} = \left[ {{{\bar s}_{k,i,j}}} \right] \in {\mathbb{C}^{{N_{{\text{RI}},k}} \times {N_{\text{L}}}}}$, ${{\mathbf{V}}_k^{\prime}} = \left[ {{{v}_{k,i,j}}} \right] \in {\mathbb{C}^{{N_{{\text{RI}},k}} \times {N_{\text{L}}}}}$, ${{\mathbf{\tilde Y}}_k} = \left[ {{{\tilde y}_{k,i,j}}} \right] \in {\mathbb{C}^{{N_{{\text{RI}},k}} \times {N_{\text{L}}}}}$ and ${{\mathbf{\hat S}}_k} = \left[ {{{\hat s}_{k,i,j}}} \right] \in {\mathbb{C}^{{N_{{\text{RI}},k}} \times {N_{\text{L}}}}}$, respectively, where ${N_{{\text{RI}},k}}{N_{\text{L}}}$ is the block length. The values of ${{\mathbf{\bar S}}_k^{\prime}}$ and ${{\mathbf{V}}_k^{\prime}}$ are updated simultaneously when the detection or decoding of ${{\mathbf{y}}_k}$ in one block is finished. Therefore, the soft output of the SISO detector can be estimated according to Bayesian Gauss-Markov theory as \cite{uchoa2016iterative,wang2006low} 
\begin{equation}
\begin{aligned} {{\mathbf{\hat s}}_{k,:,l}} = &{\mathbf{\Omega }}_k^{ - 1}{\left( {{\mathbf{\tilde H}}_k^{\text{H}}{\mathbf{\Sigma }}_k^{ - 1}{{{\mathbf{\tilde H}}}_k}{{\mathbf{V}}_{k,l}^{\prime}} + {{\mathbf{I}}_{{N_{{\text{RI}},k}}}}} \right)^{ - 1}}\\ & \cdot{\mathbf{\tilde H}}_k^{\text{H}}{\mathbf{\Sigma }}_k^{ - 1}\left( {{{{\mathbf{\tilde y}}}_{k,:,l}}}{- {{{\mathbf{\tilde H}}}_k}{{{\mathbf{\bar s}}}_{k,:,l}^{\prime}}} \right) + {{\mathbf{\bar s}}_{k,:,l}^{\prime}}, 
\end{aligned}
\label{formula2}
\end{equation}
where $l = 1,2, \ldots {N_{\text{L}}}$, ${{\mathbf{V}}_{k,l}^{\prime}}{\text{ = diag}}\left( {{{\mathbf{v}}_{k,:,l}^{\prime}}} \right)$. 
${{\mathbf{\bar s}}_{k,:,l}^{\prime}}$, ${{{\mathbf{v}}_{k,:,l}^{\prime}}}$, ${{{\mathbf{\tilde y}}}_{k,:,l}}$ and ${{\mathbf{\hat s}}_{k,:,l}}$ represent the $l$th column of ${{\mathbf{\bar S}}_k^{\prime}}$, ${{\mathbf{V}}_k^{\prime}}$, ${{\mathbf{\tilde Y}}_k}$ and ${{\mathbf{\hat S}}_k}$, respectively. 
The unbiased estimation coefficient ${{\mathbf{\Omega }}_k}$ is defined as ${{\mathbf{\Omega }}_k} \triangleq {\text{diag}}\left( {{{\boldsymbol{\rho}}_k}} \right)$ where ${{\boldsymbol{\rho}}_k} = {\left[ {{\rho _{k,1}}, \ldots ,{\rho _{k,j}}, \ldots ,{\rho _{k,{N_{{\text{RI}},k}}}}} \right]^{\text{T}}}$ and it can be calculated by 
\begin{equation}
{\rho _{k,j}} = {\mathbf{e}}_j^{\text{H}}{\left( {{\mathbf{\tilde H}}_k^{\text{H}}{\mathbf{\Sigma }}_k^{ - 1}{{{\mathbf{\tilde H}}}_k}{{\mathbf{V}}_{k,l}^{\prime}} + {{\mathbf{I}}_{{N_{{\text{RI}},k}}}}} \right)^{ - 1}}{\mathbf{\tilde H}}_k^{\text{H}}{\mathbf{\Sigma }}_k^{ - 1}{{\mathbf{\tilde H}}_k}{{\mathbf{e}}_j}.
\label{Omega3}
\end{equation}

When the detector performs iterative detection as expressed in (\ref{formula2}), the inverse operation of ${N_{{\text{RI}},k}} \times {N_{{\text{RI}},k}}$ matrix need to be performed ${N_{\text{L}}}$ times within a block. In this case, the computation complexity is $O\left( {N_{{\text{RI}},k}^3} \right) \cdot {N_{\text{L}}}$, which is extremely high. 

In order to reduce the complexity, we propose the EVD-MMSE-ISDIC algorithm. Firstly, ${\nu _{k,l}}$ is denoted as the mean of the diagonal elements of ${{\mathbf{V}}_{k,l}^{\prime}}$. Thus, ${{\mathbf{V}}_{k,l}^{\prime}}$ can be approximated as ${{\mathbf{V}}_{k,l}^{\prime}} = {\nu _{k,l}}{{\mathbf{I}}_{{N_{{\text{RI}},k}}}}$. It has been proved in \cite{tuchler2002minimum} that such an approximation does not result in much performance degradation while simplifying the calculation. Then, applying EVD, ${\mathbf{\tilde H}}_k^{\text{H}}{\mathbf{\Sigma }}_k^{ - 1}{{\mathbf{\tilde H}}_k}$ can be factorized as ${\mathbf{\tilde H}}_k^{\text{H}}{\mathbf{\Sigma }}_k^{ - 1}{{\mathbf{\tilde H}}_k}{\text{ = }}{{\mathbf{Q}}_k}{{\mathbf{\Lambda }}_k}{\mathbf{Q}}_k^{ - 1}$. 
Therefore, we obtain
\begin{equation}
{\left( {{\mathbf{\tilde H}}_k^{\text{H}}{\mathbf{\Sigma }}_k^{ - 1}{{{\mathbf{\tilde H}}}_k}{{\mathbf{V}}_{k,l}^{\prime}} + {{\mathbf{I}}_{{N_{{\text{RI}},k}}}}} \right)^{ - 1}}{\text{ = }}{{\mathbf{Q}}_k}{\left( {{\nu _{k,l}}{{\mathbf{\Lambda }}_k} + {{\mathbf{I}}_{{N_{{\text{RI}},k}}}}} \right)^{ - 1}}{\mathbf{Q}}_k^{ - 1}.
\label{Q formula}
\end{equation}
In this way, the inverse operation of ${N_{{\text{RI}},k}} \times {N_{{\text{RI}},k}}$ matrix can only be performed for one time. The simplified computation complexity is reduced to $O\left( {{N_{{\text{RI}},k}}^3} \right){\text{ + }}O\left( {{N_{{\text{RI}},k}}} \right) \cdot {N_{\text{L}}}$. Based on (\ref{Q formula}), (\ref{formula2}) can be rewritten as
\begin{equation}
\begin{aligned} 
{{\mathbf{\hat s}}_{k,:,l}} = &{\mathbf{\Omega }}_k^{ - 1}{{\mathbf{Q}}_k}{\left( {{\nu _{k,l}}{{\mathbf{\Lambda }}_k} + {{\mathbf{I}}_{{N_{{\text{RI}},k}}}}} \right)^{ - 1}}{\mathbf{Q}}_k^{ - 1}\\ & \cdot{\mathbf{\tilde H}}_k^{\text{H}}{\mathbf{\Sigma }}_k^{ - 1}\left( {{{{\mathbf{\tilde y}}}_{k,:,l}}} - {{{{\mathbf{\tilde H}}}_k}{{{\mathbf{\bar s}}}_{k,:,l}^{\prime}}} \right) + {{\mathbf{\bar s}}_{k,:,l}^{\prime}}.
\end{aligned} 
\label{final formula}
\end{equation}

\begin{figure}
\centerline{\includegraphics[scale=0.33]{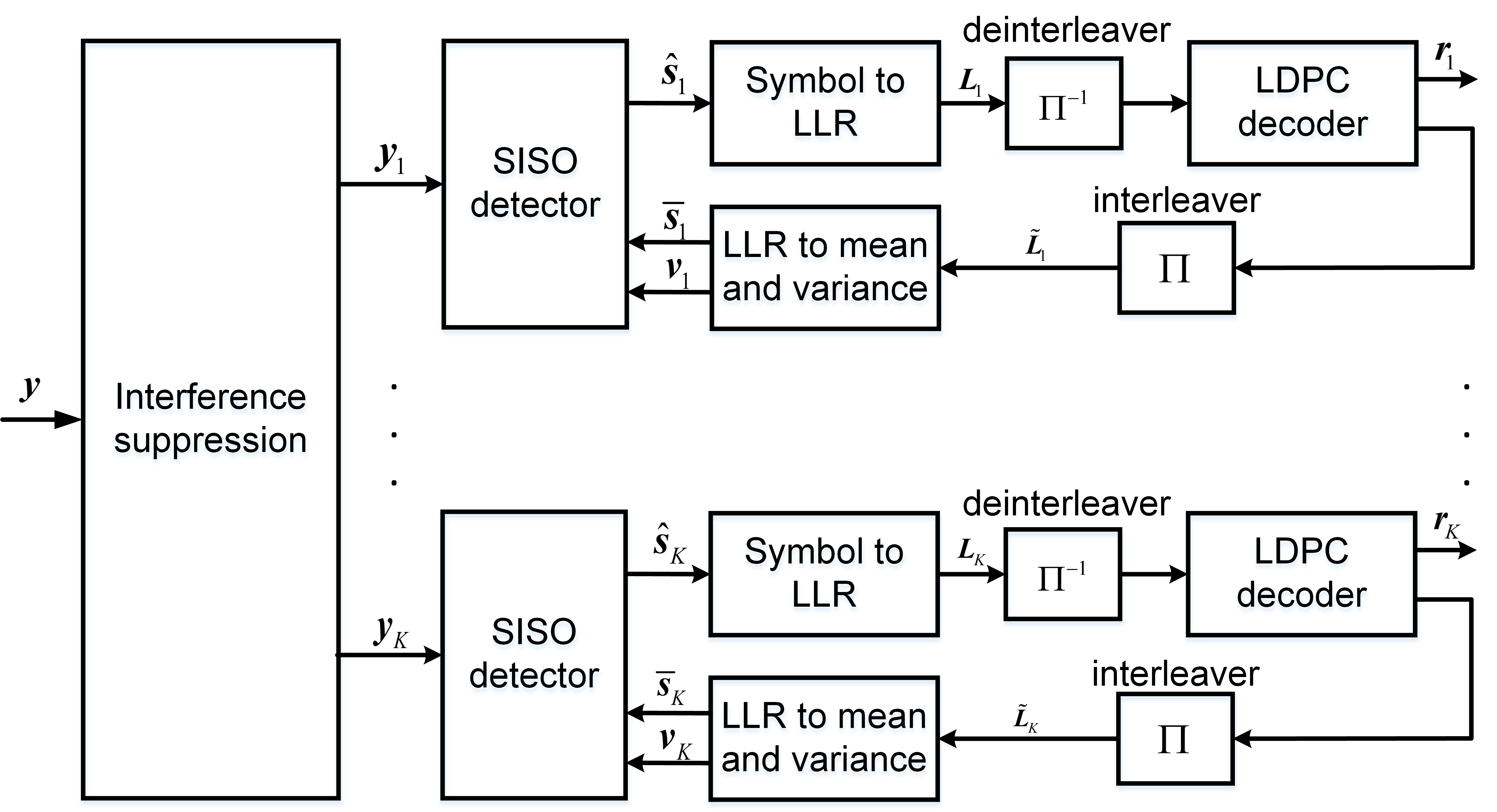}}
\caption{Receiver structure of IDD scheme.}
\label{IDD}
\end{figure}

\begin{figure}
\centerline{\includegraphics[scale=0.33]{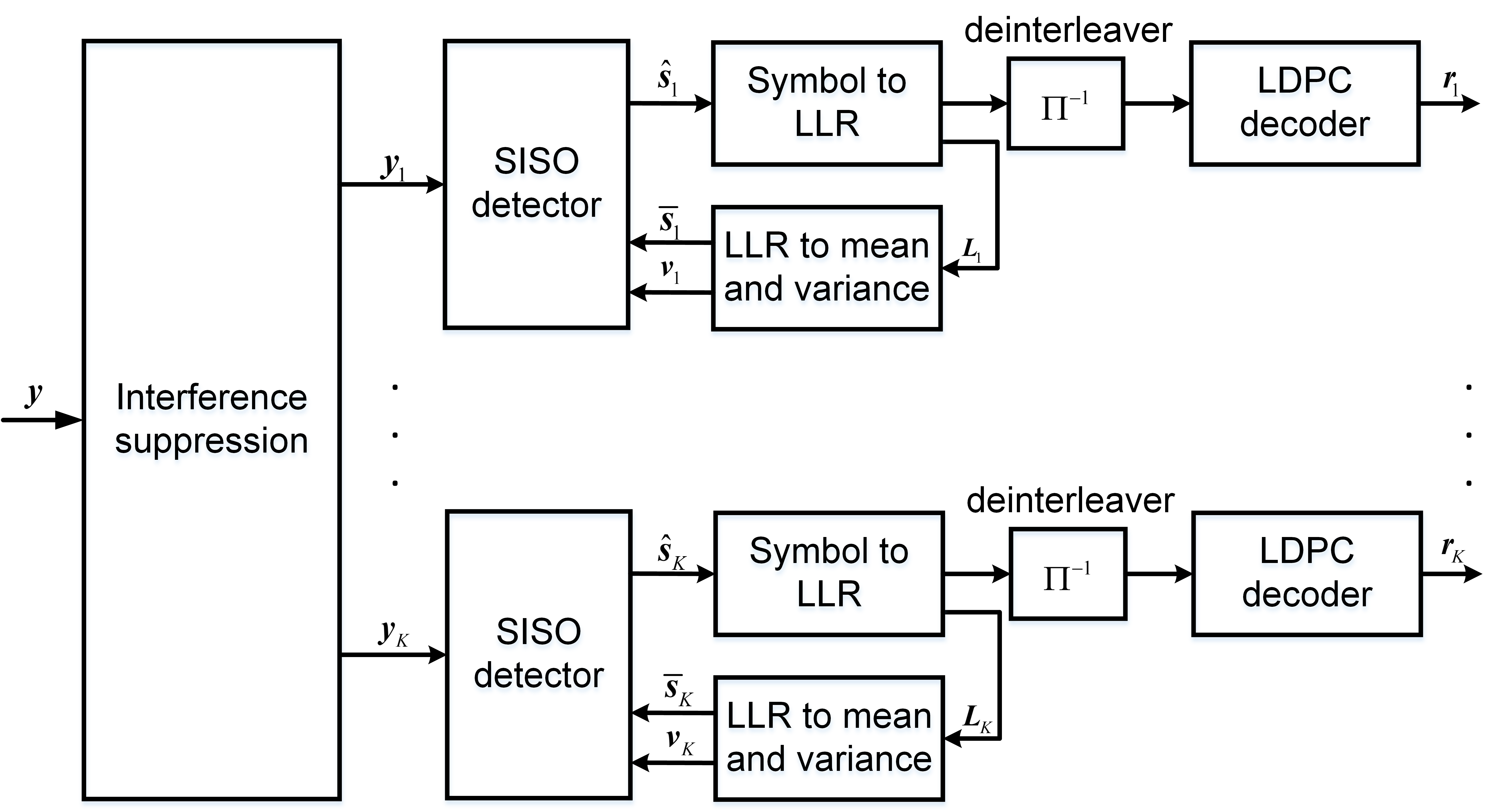}}
\caption{Receiver structure of ID scheme.}
\label{ID}
\end{figure}

Two iteration schemes named as IDD and ID are given based on the low-complexity EVD-MMSE-ISDIC to improve the receiving performance of large-scale D-MIMO systems. The receiver structures of IDD scheme and ID scheme are shown in Fig. \ref{IDD} and Fig. \ref{ID}, respectively. In the IDD scheme, interleaving/deinterleaving is first performed. Then, the low-density parity-check (LDPC) decoder generates extrinsic LLRs by using the correlation between different code structures introduced by the encoder. The extrinsic LLRs are used as the \emph{a priori} information to calculate ${{\mathbf{\bar s}}_k}$ and ${{\mathbf{v}}_k}$ and returned to the detector afterwards. The main difference between IDD scheme and ID scheme is that in the ID scheme, the extrinsic LLRs calculated by the output of the detector is directly used as the \emph{a priori} information to calculate ${{\mathbf{\bar s}}_k}$ and ${{\mathbf{v}}_k}$. Then, they are returned to the detector directly.  

In the first iteration, we set ${{\mathbf{\bar s}}_{k,:,l}^{\prime}} = {\bf{0}}$ and ${{\mathbf{V}}_{k,l}^{\prime}}{\rm{ = }}{{\bf{I}}_{{N_{{\rm{RI}},k}}}}$. According to matrix inversion lemma 
\[{\left( {{\mathbf{A}} + {\mathbf{BCD}}} \right)^{ - 1}} = {{\mathbf{A}}^{ - 1}} - {{\mathbf{A}}^{ - 1}}{\mathbf{B}}{\left( {{\mathbf{D}}{{\mathbf{A}}^{ - 1}}{\mathbf{B}} + {{\mathbf{C}}^{ - 1}}} \right)^{ - 1}}{\mathbf{D}}{{\mathbf{A}}^{ - 1}},\] 
(\ref{final formula}) is simplified to
\begin{equation}
{{\bf{\hat s}}_{k,:,l}}{\rm{ = }}{\left( {{\bf{\tilde H}}_k^{\rm{H}}{\bf{\Sigma }}_k^{ - 1}{{{\bf{\tilde H}}}_k} + {{\bf{I}}_{{N_{{\rm{RI}},k}}}}} \right)^{ - 1}}{\bf{\tilde H}}_k^{\rm{H}}{\bf{\Sigma }}_k^{ - 1}{{\bf{\tilde y}}_{k,:,l}}\label{formula7},
\end{equation}
which is the expression of the usual linear MMSE (LMMSE) detection algorithm.
In the last iteration, the decoders produce hard decisions on information bits in the last iteration.
The proposed EVD-MMSE-ISDIC detection algorithm is described in Algorithm 1. 

\begin{algorithm}
\label{algorithm1}
\caption{EVD-MMSE-ISDIC Detection Algorithm}
\begin{algorithmic}[1]
\STATE \textbf{Require} Iterative Scheme ($IS$), ${\mathbf{y}}$, ${{\mathbf{H}}_{k}}$, Number of Iterations (${N_{{\text{I}}}}$).
\STATE \textbf{Set} ${{\mathbf{\bar s}}_k}{\text{ = zeros}}\left( {{N_{{\text{RI}},k}}{N_{\text{L}}},1} \right)$, ${{\mathbf{v}}_k}{\text{ = ones}}\left( {{N_{{\text{RI}},k}}{N_{\text{L}}},1} \right)$.
\STATE Calculate ${{\mathbf{\Sigma }}_k}$ with equation (\ref{formula1}) and calculate ${\mathbf{\tilde H}}_k^{\text{H}}{\mathbf{\Sigma }}_k^{ - 1}{{\mathbf{\tilde H}}_k} = {\mathbf{H}}_{k}^{\text{H}}{\mathbf{W}}_{{\text{IS}},k}^{\text{H}}{\mathbf{\Sigma }}_k^{ - 1}{{\mathbf{W}}_{{\text{IS}},k}}{{\mathbf{H}}_{k}}$.
\STATE \textbf{Detection Algorithm—EVD-MMSE-ISDIC}
\STATE ${\text{EVD}}\left( {{\mathbf{\tilde H}}_k^{\text{H}}{\mathbf{\Sigma }}_k^{ - 1}{{{\mathbf{\tilde H}}}_k}} \right)$: ${\mathbf{\tilde H}}_k^{\text{H}}{\mathbf{\Sigma }}_k^{ - 1}{{\mathbf{\tilde H}}_k}{\text{ = }}{{\mathbf{Q}}_k}{{\mathbf{\Lambda }}_k}{\mathbf{Q}}_k^{ - 1}$.
\FOR{$it = 1 \to {N_{{\text{I}}}}$}
\FOR{$l = {\text{ }}1 \to {N_L}$}
\STATE Calculate ${\left( {{\nu _{k,l}}{{\mathbf{\Lambda }}_k} + {\mathbf{I}}} \right)^{ - 1}}$ , ${{\mathbf{Q}}_k}{\left( {{\nu _{k,l}}{{\mathbf{\Lambda }}_k} + {\mathbf{I}}} \right)^{ - 1}}{\mathbf{Q}}_k^{ - 1}$.
\STATE Calculate ${{\mathbf{\Omega }}_k}$ with equations (\ref{Omega3}). 
\STATE Calculate ${{\mathbf{\hat s}}_{k}}$ with equations (\ref{final formula}).
\ENDFOR
\STATE Soft demodulation and obtain the extrinsic LLRs ${{\mathbf{l}}_k}$.
\IF{$IS = {\text{IDD}}$}
\STATE \textbf{LDPC Decoding}
\STATE Obtain the output ${{\mathbf{r}}_k}$ and the extrinsic LLRs ${{\mathbf{\tilde l}}_k}$. 
\ENDIF
\STATE Calculate ${{\mathbf{\bar s}}_k}$ and ${{\mathbf{v}}_k}$ with equations from (\ref{formula3}) up to (\ref{formula6}). 
\ENDFOR
\IF{$IS = {\text{ID}}$}
\STATE \textbf{LDPC Decoding}
\STATE Obtain the output ${{\mathbf{r}}_k}$.
\ENDIF
\end{algorithmic}
\end{algorithm}

\begin{remark}
It should be noted that the proposed EVD-MMSE-ISDIC detection algorithm is also applicable to the downlink receiver. The detection expression in the downlink transmission is the same as (\ref{final formula}) without ${\mathbf{W}}_{{\text{IS}},k}$ and ${\mathbf{W}}_{{\text{IS}},k}^{\text{H}}$. 
\end{remark}

\subsection{Calculation of Symbol Mean and Variance}\label{Calculation of Symbol Mean and Variance}
In this subsection, we discuss the method of calculating the mean and variance of ${{\mathbf{\hat s}}_k}$ from the \emph{a priori} LLRs.
The LLR of ${s_{k,j}}$ is denoted as ${{\mathbf{l}}_{k,j}}{\text{ = }}{\left[ {{L_{k,j,1}}, \ldots ,{L_{k,j,{M_{\text{c}}}}}} \right]^{\text{T}}}$. Thus, the probability of the symbol ${s_{k,j}}$ can be calculated by \cite{wang2004low}
\begin{equation}
P\left[ {{s_{k,j}} = \alpha \left( {\mathbf{d}} \right)} \right]{\text{ = }}\prod\limits_{i = 1}^{{M_{\text{c}}}} {\left[ {\frac{1}{{1 + \exp \left( { - {{\tilde d}_i}{L_{k,j,i}}} \right)}}} \right]}, 
\label{formula3}
\end{equation}
where ${M_c}$ is the bit numbers contained in a constellation symbol, ${\mathbf{d}}$ is a ${M_{\text{c}}} \times 1$ vector of bits, $\alpha \left( {\mathbf{d}} \right)$ represents the constellation symbol of ${\mathbf{d}}$ and the elements in ${M_c} \times 1$ vector ${\mathbf{\tilde d}}$ are defined as
\begin{equation}{\tilde d_i} = \left\{ \begin{gathered}
   + 1,{\text{     }}{d_i} = 1 \hfill \\
   - 1,{\text{     }}{d_i} = 0 \hfill \\ 
\end{gathered}  \right..
\label{formula4}
\end{equation}
The mean and variance of ${{\mathbf{s}}_k}$ can be calculated by \cite{wang2004low,wang1999iterative}
\begin{equation}
{\bar s_{k,j}} = \sum\limits_{{\mathbf{d}} \in \Theta }^{} {\alpha \left( {\mathbf{d}} \right)} P\left[ {{s_{k,j}} = \alpha \left( {\mathbf{d}} \right)} \right],
\label{formula5}
\end{equation}
\begin{equation}
{v_{k,j}} = {\sum\limits_{{\mathbf{d}} \in \Theta }^{} {\left| {\alpha \left( {\mathbf{d}} \right)} \right|} ^2}P\left[ {{s_{k,j}} = \alpha \left( {\mathbf{d}} \right)} \right] - {\left| {{{\bar s}_{k,j}}} \right|^2}.
\label{formula6}
\end{equation}
\begin{remark}
In practical situations, in order to further reduce the computational complexity, the values of $\frac{1}{{1 + \exp \left( x \right)}}$ are mapped to a hash table, where $x$ ranges from $-\max \left( {{L_{k,j,i}}} \right)$ to $\max \left( {{L_{k,j,i}}} \right)$. Therefore, we can obtain the values directly according to index ${{L_{k,j,i}}}$ when calculating (\ref{formula3}). 
\end{remark}

\section{Large-scale D-MIMO prototyping system} \label{Large-scale D-MIMO prototyping system}
The large-scale D-MIMO prototyping system consists of $M$ RAUs and $K$ users (up to 16 RAUs and 16 users), each RAU and user is equipped with 8 antennas. The carrier frequency of the system is 3.5 GHz and the bandwidth is 96 MHz. We consider a time-division-duplex (TDD) system with a 1:1 ratio between uplink and downlink. The baseband processing is performed by base station-baseband processing units (BS-BPUs) and user equipment-BPUs (UE-BPUs). Each BPU has an Intel Xeon E5 72 core general-purpose processor (GPP) based on Sandy Bridge architecture. Since the demand for computing capability of the RAU side is much larger than that of the user side, the RAUs are connected to the cloud processing center composed of large-scale high-performance processors. That is also in line with the practical scenario where the RAUs have powerful computing capability while the computing performance of the terminal device is limited. Tab. \ref{Parameters} lists the main parameters of the system.

\begin{figure}
\centerline{\includegraphics[scale=0.3]{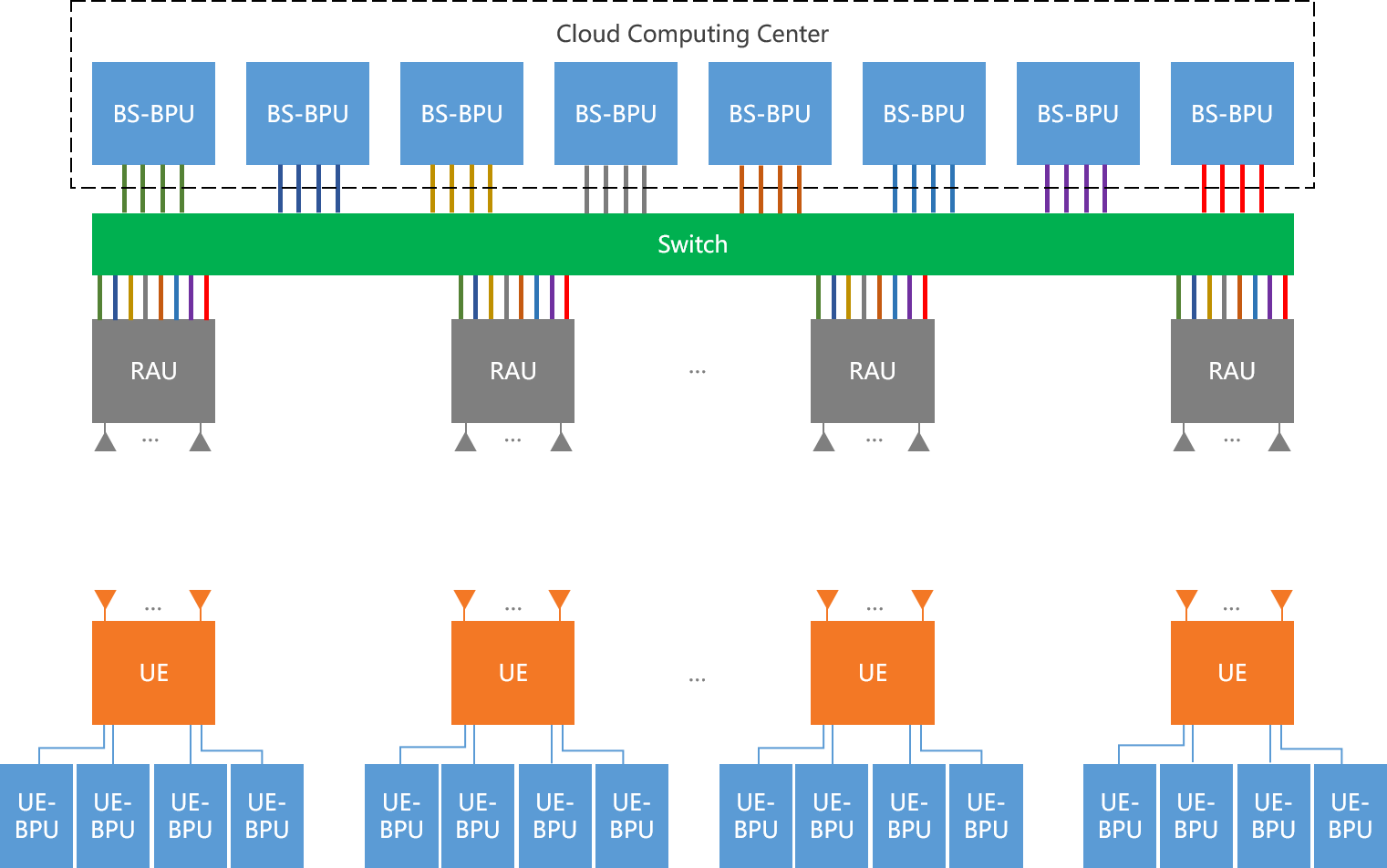}}
\caption{System architecture of a large-scale D-MIMO prototyping system with $M = 4$, $K = 4$, ${P_{\text{B}}} = 8$ and ${P_{\text{U}}} = 4$. Specific signals are transmitted to specific BS-BPU for processing, e.g., all red sub-band signals received by all RAUs are transmitted to the rightmost BS-BPU.}
\label{System architecture}
\end{figure}

\begin{table}
\caption{Parameters of large-scale D-MIMO prototyping system}
\begin{center}
\begin{tabular}{|c|c|}
\hline
\textbf{Parameter}&\textbf{Value} \\
\hline
Subcarrier space ${\Delta _{\text{F}}}$ & 30 KHz \\
\hline
Total number of subcarriers, IFFT/FFT size ${N_{{\text{FFT}}}}$ & 4096 \\
\hline
Sampling rate ${f_{\text{s}}}$ & 122.88 MHz \\
\hline
Number of used subcarriers ${N_{\text{D}}}$ & 3208 \\
\hline
Bandwidth ${W}$ & 96 MHz \\
\hline
IFFT/FFT cycle ${T_{{\text{FFT}}}} = \frac{1}{{{\Delta _{\text{F}}}}}$ & 33.333 $\mu$s \\
\hline
Cyclic prefix length ${T_{{\text{CP}}}}$, ${N_{{\text{CP}}}}$ & 4.167 $\mu$s, 512 \\
\hline
OFDM symbol duration ${T_{{\text{SYM}}}} = {T_{{\text{CP}}}} + {T_{{\text{FFT}}}}$ & 37.5 $\mu$s \\
\hline
\end{tabular}
\label{Parameters}
\end{center}
\end{table}
System architecture of a large-scale D-MIMO prototyping system is shown in Fig. \ref{System architecture}. There are ${P_{\text{B}}}$ BS-BPUs on the RAU side and the 96 MHz bandwidth of each RAU is divided into ${P_{\text{B}}}$ parts. The specific sub-band is transmitted to the specific BS-BPU in the cloud computing center through switches. On the user side, each user is equipped with ${P_{\text{U}}}$ UE-BPUs, and each UE-BPU processes signals with $\frac{{96}}{{{P_{\text{U}}}}}$ MHz bandwidth.

\begin{figure*}
\centerline{\includegraphics[scale=0.2]{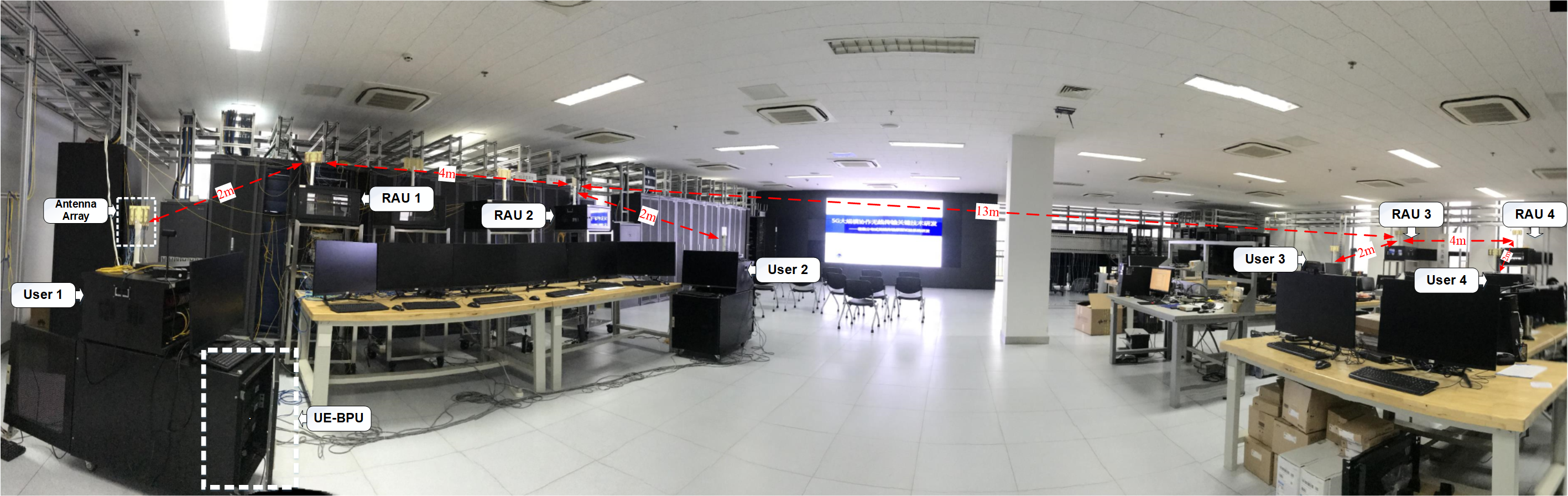}}
\caption{A 32$ \times $32 layout of the large-scale D-MIMO prototyping system.}
\label{actual layout}
\end{figure*}
According to the channel quality indication (CQI), the users obtain the number of data streams ${N_{{\text{RI}}}}$, the modulation order ${Q_{\text{m}}}$, and the coding rate ${R_{\text{c}}}$. For each user, the data generated by UE-BPUs is transmitted to the radio frequency (RF) electrical control panel after encoding, interleaving, modulating, subcarrier mapping, antenna mapping and OFDMA. On the RAU side, the receiver performs interference suppression, MIMO detecting, demodulating, deinterleaving and decoding orderly to obtain the transmitted data.

To improve system performance, we use Intel's math kernel library (MKL) to perform baseband processing at BPUs, which provides a number of highly optimized routines that support vector-vector, matrix-vector and matrix-matrix operations for real and complex single and double precision data. In addition, through multi-core parallel programming, the computational efficiency of the processor is greatly improved. For example, in multi-user scenarios, one core of a BS-BPU processes one sub-band of a user, including signal receiving and transmitting. After the interference suppression matrix is multiplied by the received signal, the multi-user receiver is decomposed into $K$ independent single-user receivers which are implemented in $K$ independent core sets.

\section{Experimental results}\label{Experimental results}
Since the complexity of the proposed EVD-MMSE-ISDIC algorithm is reduced dramatically, both IDD and ID schemes can be implemented in prototyping systems. Therefore, experimental validation of EVD-MMSE-ISDIC is performed in the large-scale D-MIMO prototyping system. Furthermore, we compare the algorithm performance of the proposed iterative algorithm with non-iterative LMMSE to demonstrate that both IDD scheme and ID scheme offer a higher-performance solution. A 32$ \times $32 system configuration (4 users and 4 RAUs, each equipped with 8 antennas) is considered. The layout of the large-scale D-MIMO prototyping system is shown in Fig. \ref{actual layout}. Each user is equipped with 4 UE-BPUs while all RAUs share 8 BS-BPUs. When CQI is less than 6, error-free transmission can be achieved in the uplink generally. Therefore, we focus on the condition when CQI equals 6 (${N_{{\text{RI}}}} = 4$, ${Q_{\text{m}}} = 4$, ${R_{\text{c}}} = 3/4$) and 7 (${N_{{\text{RI}}}} = 4$, ${Q_{\text{m}}} = 6$, ${R_{\text{c}}} = 2/3$). Without considering the pilot overhead, the uplink spectral efficiency of the system can reach 48 bps/Hz when CQI equals 6 and 64 bps/Hz when CQI equals 7.

\begin{figure}
\centerline{\includegraphics[scale=0.6]{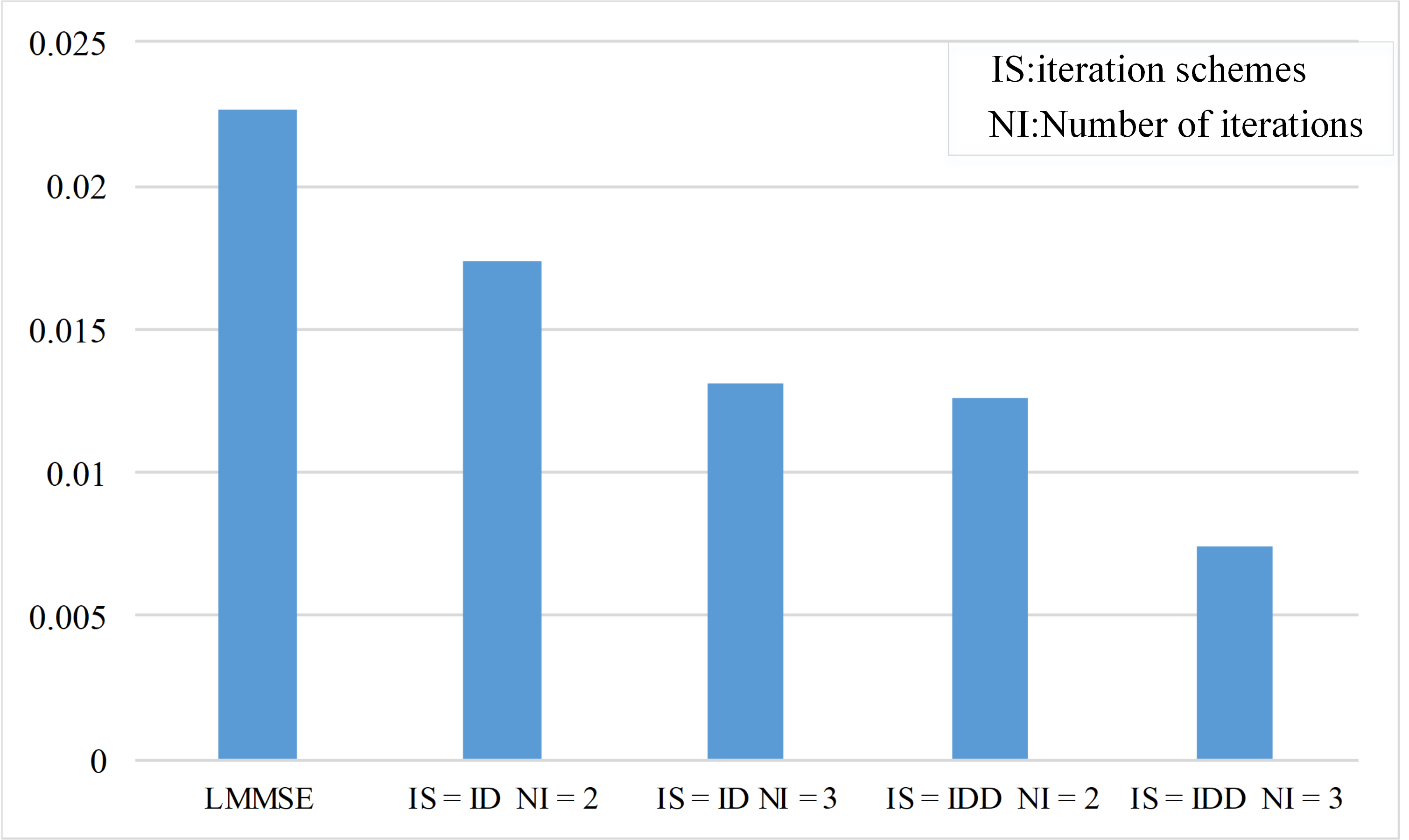}}
\caption{The average number of error blocks in the uplink with different iteration schemes
and different iteration numbers when CQI = 6.}
\label{CQI6}
\end{figure}

\begin{figure}
\centerline{\includegraphics[scale=0.6]{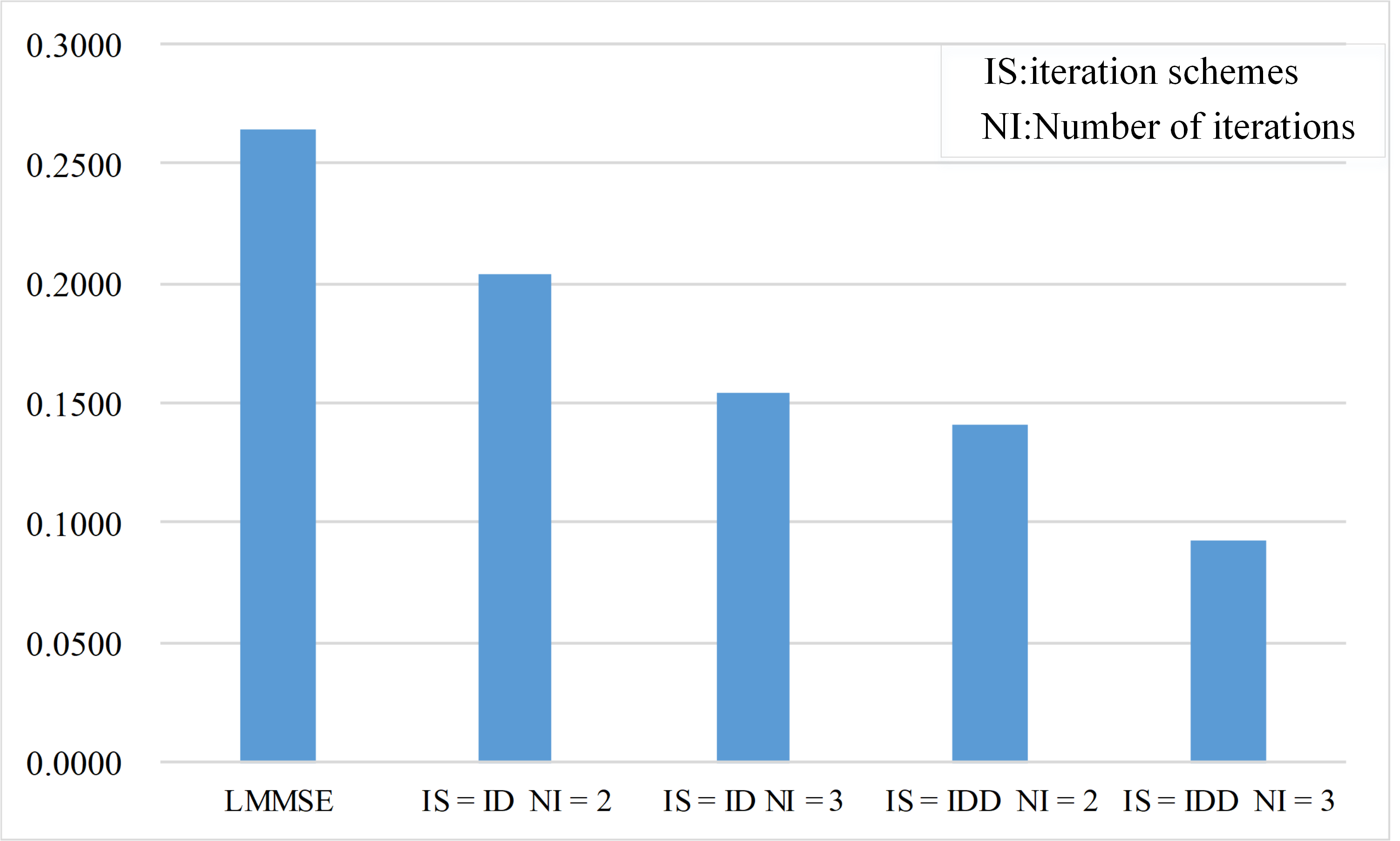}}
\caption{The average number of error blocks in the uplink with different iteration schemes
and different iteration numbers when CQI = 7.}
\label{CQI7}
\end{figure}

The statistical averaging on the number of error blocks within a time period is performed. The average number of error blocks of all BS-BPUs with different iteration schemes and different number of iterations are recorded and shown in Fig. \ref{CQI6} and Fig. \ref{CQI7}. Considering the impact of system delay, the maximum number of iteration is set to be three. The experimental results demonstrate that the proposed EVD-MMSE-ISDIC algorithm can greatly reduce the average number of error blocks compared with the conventional non-iterative LMMSE receiver. 

Compared with the LMMSE receiver, in the ID scheme, the average number of error blocks can be reduced by 23\% and 40\% with two iterations and three iterations, respectively. Besides, in the IDD scheme, a approximately 45\% and 64\% decrease for the average number of blocks in comparison with LMMSE can be achieved with two iterations and three iterations, respectively. Furthermore, the IDD algorithm converges faster and has better receiving performance compared with ID. However, the decoding operation need to be performed for multiple times in IDD and higher calculation capability of the system is required. In the case when the computing capability of the practical system is limited, ID scheme with three iterations or IDD scheme with two iterations are preferred since they are cost-effective.
\section{Conclusion}\label{Conclusion}
In this paper, we have presented a low-complexity iterative SISO detection algorithm in a large-scale distributed MIMO system. The uplink interference suppression matrix was designed to decompose the received multi-user signal into independent single-user receptions. Then, EVD-MMSE-ISDIC algorithm was proposed to perform low-complexity detection of the decomposed signals. Based on EVD-MMSE-ISDIC, two different schemes, named as IDD scheme and ID scheme, have been given and the proper detection scheme can be chosen according to the computing capability of the practical system. A large-scale distributed MIMO prototyping system has been introduced to perform experimental validation of the proposed detection schemes. Experimental results demonstrate that the proposed iterative receiver can greatly reduce the average number of error blocks of the system. Compared with the non-iterative LMMSE receiver, the average number of error blocks can be reduced by 64\% and 40\% with three iterations in IDD and ID scheme, respectively.

\section*{Acknowledgment}
This work was supported in part by National Natural Science Foundation of China (NSFC) (Grant NO. 61871122, 61801168), National Key Special Program No.2018ZX03001008-002, and six talent peaks project in Jiangsu province.

\renewcommand\refname{Reference}
\bibliographystyle{ieeetr}
\bibliography{test}

\begin{thebibliography}{10}

\bibitem{joung2014energy}
J.~Joung, Y.~K. Chia, and S.~Sun, ``Energy-efficient, large-scale
  distributed-antenna system ({L-DAS}) for multiple users,'' {\em IEEE Journal
  of Selected Topics in Signal Processing}, vol.~8, no.~5, pp.~954--965, 2014.

\bibitem{you2010cooperative}
X.-H. You, D.-M. Wang, B.~Sheng, X.-Q. Gao, X.-S. Zhao, and M.~Chen,
  ``Cooperative distributed antenna systems for mobile communications,'' {\em
  IEEE Wireless Communications}, vol.~17, no.~3, p.~35, 2010.

\bibitem{almelah2017spectral}
H.~B. Almelah and K.~A. Hamdi, ``Spectral efficiency of distributed large-scale
  {MIMO} systems with {ZF} receivers,'' {\em IEEE Transactions on Vehicular
  Technology}, vol.~66, no.~6, pp.~4834--4844, 2017.

\bibitem{wang2013spectral}
D.~Wang, J.~Wang, X.~You, Y.~Wang, M.~Chen, and X.~Hou, ``Spectral efficiency
  of distributed {MIMO} systems,'' {\em IEEE Journal on Selected Areas in
  Communications}, vol.~31, no.~10, pp.~2112--2127, 2013.

\bibitem{elghariani2016low}
A.~Elghariani and M.~Zoltowski, ``Low complexity detection algorithms in
  large-scale {MIMO} systems,'' {\em IEEE Transactions on Wireless
  Communications}, vol.~15, no.~3, pp.~1689--1702, 2016.

\bibitem{rusek2013scaling}
F.~Rusek, D.~Persson, B.~K. Lau, E.~G. Larsson, T.~L. Marzetta, O.~Edfors, and
  F.~Tufvesson, ``Scaling up {MIMO}: Opportunities and challenges with very
  large arrays,'' {\em IEEE signal processing magazine}, vol.~30, no.~1,
  pp.~40--60, 2013.

\bibitem{rossler2003mmse}
J.~F. Ro{\ss}ler, W.~H. Gerstacker, and J.~B. Huber, ``{MMSE-based} iterative
  equalization with soft feedback for {QAM} transmission over sparse
  channels,'' in {\em Telecommunications, 2003. ICT 2003. 10th International
  Conference on}, vol.~1, pp.~566--571, IEEE, 2003.

\bibitem{cao2007relation}
F.~Cao, J.~Li, and J.~Yang, ``On the relation between {PDA} and {MMSE-ISDIC},''
  {\em IEEE Signal Processing Letters}, vol.~14, no.~9, pp.~597--600, 2007.

\bibitem{uchoa2016iterative}
A.~G. Uchoa, C.~T. Healy, and R.~C. de~Lamare, ``Iterative detection and
  decoding algorithms for {MIMO} systems in block-fading channels using {LDPC}
  codes,'' {\em IEEE Transactions on Vehicular Technology}, vol.~65, no.~4,
  pp.~2735--2741, 2016.

\bibitem{wang2006low}
D.~Wang, X.~Gao, and X.~You, ``Low complexity turbo receiver for multi-user
  {STBC} block transmission systems,'' {\em IEEE transactions on wireless
  communications}, vol.~5, no.~10, pp.~2625--2632, 2006.

\bibitem{tuchler2002minimum}
M.~Tuchler, A.~C. Singer, and R.~Koetter, ``Minimum mean squared error
  equalization using a priori information,'' {\em IEEE Transactions on Signal
  processing}, vol.~50, no.~3, pp.~673--683, 2002.

\bibitem{wang2004low}
D.~Wang, J.~Hua, X.~Gao, X.~You, and W.~Park, ``Low complexity iterative
  receiver for multiuser {STBC} block transmission systems,'' in {\em Global
  Telecommunications Conference, 2004. GLOBECOM'04. IEEE}, vol.~2,
  pp.~1086--1090, IEEE, 2004.

\bibitem{wang1999iterative}
X.~Wang and H.~V. Poor, ``Iterative (turbo) soft interference cancellation and
  decoding for coded {CDMA},'' {\em IEEE Transactions on communications},
  vol.~47, no.~7, pp.~1046--1061, 1999.

\end{thebibliography}
\end{document}